\documentclass[12pt,a4paper,aps,tightenlines,nofootinbib,showkeys,superscriptaddress]{revtex4-1}
\pdfoutput=1
\usepackage{fullpage}
\usepackage{float}
\usepackage{amsfonts}
\usepackage{amsmath}
\usepackage{slashed}
\usepackage{mathrsfs}
\usepackage[scr=rsfs,cal=boondox]{mathalfa}
\usepackage{amssymb}
\usepackage{graphicx}
\usepackage{makeidx}
\usepackage{cancel}
\usepackage{eepic}
\usepackage{epsfig}
\usepackage{latexsym}
\usepackage{mathtools}
\usepackage[dvipsnames]{xcolor}
\usepackage{float}
\usepackage{multirow}
\usepackage{siunitx}
\usepackage[export]{adjustbox}
\usepackage{xurl,hyperref}
\usepackage{enumitem}
\usepackage{soul}
\hypersetup{colorlinks=true,citecolor=red,linkcolor=NavyBlue,urlcolor=NavyBlue}
\usepackage[utf8]{inputenc}
\usepackage[caption=false]{subfig}
 
\usepackage{natbib}
\usepackage{mathptmx}

\def\ba{\begin{array}}       
	\def\ea{\end{array}}

\def\beq{\begin{eqnarray}}
\def\eeq{\end{eqnarray}}
\begin{document}
	\title{Next-to-leading order QCD corrections to $Z\to q\bar{q}\gamma$, $q\bar{q}\gamma\gamma$}
	\author{Pankaj Agrawal}
\email{pankaj.agrawal@tcgcrest.org}
\affiliation{Centre for Quantum Engineering, Research and Education, TCG CREST, Kolkata, India}
\author{Subhadip Bisal}
\email{subhadip.b@iopb.res.in}
\affiliation{Institute of Physics, Sachivalaya Marg, Bhubaneswar 751 005, India}
\affiliation{Homi Bhabha National Institute, Training School Complex, Anushakti Nagar, Mumbai 400 094, India}
	\author{Biswajit Das}
\email{biswajitd@imsc.res.in}
\affiliation{The Institute of Mathematical Sciences, Chennai, Tamil Nadu 600113, India}
\author{Debottam Das}
\email{debottam@iopb.res.in}
\affiliation{Institute of Physics, Sachivalaya Marg, Bhubaneswar 751 005, India}
\affiliation{Homi Bhabha National Institute, Training School Complex, Anushakti Nagar, Mumbai 400 094, India}

	\date{\today}
	
	\begin{abstract}
		\noindent

We consider the rare decay channels of the $Z$ boson: $Z \to \text{two}\  \textrm{jets} + \gamma$ and $Z \to \text{two}\ \textrm{jets} +2\, \gamma$. To obtain the widths and distributions for these processes, we compute the effect of NLO QCD corrections to the processes  $Z \to q {\bar q}+ \gamma$ and $Z \to q {\bar q} +2\, \gamma$. We find that these corrections reduce the widths of these processes by about $6.03\%$ and $12.39\%$, respectively. The reduction in the partial widths is larger at the jet level. These NLO-improved decay observables may be tested in future runs of the LHC or at future $e^{+}e^{-}$ colliders.

	\end{abstract}
	
	\maketitle
	\section{Introduction}
	
	Theoretical calculations of electroweak precision tests of the
        Standard Model (SM) align closely with data from the
        Large Hadron Collider (LHC)~\cite{ParticleDataGroup:2022pth}.
        In the electroweak sector, various $Z$-pole measurements have
        already been tested at SLC and LEP~\cite{ALEPH:2005ab}. 
        A more precise determination of the
        $Z$ boson properties associated with radiative corrections at future $e^+e^-$ machines
        will not only serve as critical tests for the SM but also, should any discrepancies with SM predictions arise during these experiments, could indicate New Physics (NP) beyond the current framework. Thus, the calculation of the $Z$
        boson observables at one-loop and beyond are necessary
        to mitigate theoretical uncertainties effectively.
        
At the colliders, electroweak (EW) observables typically include the $Z$ boson mass $M_Z$, total and partial widths of the $Z$ boson and the
        $Z$ boson couplings.
        In particular, the partial width ($\Gamma_{Z_{ij..}}$)
        ($ij..$ refer to the different SM final states) or total width ($\Gamma_Z$) can be ascribed to an
        important property. In the SM, $Z$ boson width is known involving the complete EW two-loop and some
leading partial QCD or mixed three-loop terms~\cite{Blondel:2018mad}.
The current theoretical uncertainty from unaccounted higher-order corrections is estimated at approximately $\delta\Gamma_Z^{(1)} \sim 0.4$~MeV~\cite{Dubovyk:2018rlg, Blondel:2018mad}. Assuming EW three-loop corrections are determined, this uncertainty could be refined to $\delta\Gamma_Z^{(2)} \sim 0.15$~MeV~\cite{Blondel:2018mad}. Furthermore, if the dominant four-loop corrections were known, an upper limit on the theoretical error could be set at $\delta\Gamma_Z^{(3)} < 0.07$~MeV~\cite{Blondel:2018mad} (or $\sim 25$~keV~\cite{Bernardi:2022hny}), although precise estimates remain challenging at this level.
These values fall below the experimental precision of LEP ($\sim$2~MeV for the mass and decay width of the $Z$ boson)~\cite{ALEPH:2013dgf, ALEPH:2005ab}. But most importantly, the first two values exceed the anticipated precision of the FCC-ee at the Tera-$Z$ stage ($\sim$0.1~MeV)~\cite{Tenchini:2014lma, ParticleDataGroup:2016lqr,Blondel:2018mad}. 
Thus, it is important to figure out rare decays, as well as tiny higher-order corrections to existing $Z$ boson decays.
        For instance, the rare $Z$ boson decays,
        $Z\to J/\Psi \gamma, \Upsilon \gamma, \phi \gamma$ offer distinctive
        opportunities~\cite{Huang:2014cxa} in terms of the discovery and the total width of $Z$ boson.\footnote{In the case
        of extended Higgs sector, $Z$ boson rare decays may become a useful tool to test the imprints of NP~\cite{Bisal:2023mgz}.} We recall that one-loop corrections to the $Z$ boson Electroweak Precision
Observables (EWPOs)
were initially documented in Ref.~\cite{Akhundov:1985fc, Bernabeu:1987me, Beenakker:1988pv}.
The two-loop
corrections to the production and decay of $Z$ boson are elucidated in
Refs.~\cite{Dubovyk:2016aqv, Dubovyk:2018rlg} (for a recent update, see Ref.~\cite{Blondel:2018mad}).
\footnote{The corrections to the Fermi constant, which can be used to predict
        the $Z$-pole parameters, are calculated at two-loop
        $\mathcal{O}(\alpha_{\rm EW} \alpha_s)$~\cite{Djouadi:1987gn, Djouadi:1987di, Kniehl:1989yc, Kniehl:1991gu, Djouadi:1993ss}
        and fermionic $\mathcal{O}(\alpha_{\rm EW}^2)$~\cite{Barbieri:1992nz, Barbieri:1992dq, Fleischer:1993ub, Fleischer:1994cb, Degrassi:1996mg, Degrassi:1996ps, Degrassi:1999jd, Freitas:2000gg, Freitas:2002ja, Awramik:2004ge, Hollik:2005va, Awramik:2008gi, Freitas:2012sy, Freitas:2013dpa, Freitas:2014hra}. The ``fermionic'' refers to diagrams with at least one closed
fermion loop. The leading three- and four-loop results, enhanced by powers of
        the top Yukawa coupling $y_t$, were obtained at order $\mathcal{O}(\alpha_t \alpha_s^2)$~\cite{Avdeev:1994db, Chetyrkin:1995ix}, $\mathcal{O}(\alpha_t^2\alpha_s)$, $\mathcal{O}(\alpha_t^3)$~\cite{vanderBij:2000cg, Faisst:2003px},
        and $\mathcal{O}(\alpha_t\alpha_s^3)$~\cite{Schroder:2005db, Chetyrkin:2006bj, Boughezal:2006xk},
        where $\alpha_t=y_t^2/4\pi$.}   

 In this article, we study the complete one-loop QCD corrections to the $Z$ boson decay processes into two jets $+\gamma$ and two jets $+2\gamma$ final states, which are, to the best of our knowledge, not available in the literature. While the decay widths of these processes reside within the $\mathcal{O}$(MeV) and $\mathcal{O}$(keV) ranges, it is imperative to incorporate these partial widths, 
 into the total decay widths of $Z$ boson. 
 For the process $Z\to$two jets $+\gamma$, we find that the partial decay width is significantly larger than the theoretical uncertainties $\delta\Gamma_Z^{(1)}$ or $\delta\Gamma_Z^{(2)}$. In contrast, for $Z\to$two jets $+2\gamma$, the partial decay width may become relevant when considering the projected uncertainty $\delta\Gamma_Z^{(3)}$. The future $e^+e^-$ colliders, including the ILC, FCC-ee, and CEPC, are designed to operate precisely at the $Z$ boson resonance (i.e., $\sqrt{s}=M_Z$)~\cite{ILC:2013jhg, TLEPDesignStudyWorkingGroup:2013myl, dEnterria:2016sca, CEPC-SPPCStudyGroup:2015csa} are planned to observe up to $5\times 10^{12}$ $Z$ boson decays. These statistics are several orders of magnitude larger than at LEP and would lead to very accurate experimental measurements of EWPOs.

The article is organized as follows. Sec~\ref{sec:processes} provides an overview of the processes under study and the corresponding Feynman diagrams. Sec.~\ref{sec:computation} details the computations of tree-level and one-loop amplitudes using the spinor helicity techniques, with a brief discussion on UV renormalization in Sec.~\ref{subsec:uv_ren} and a review of IR singularities and dipole subtraction techniques in Sec.~\ref{subsec:ir_ds}. Sec.~\ref{subsec:checks} outlines the consistency checks conducted to validate our computations. The numerical results are presented in Sec.~\ref{sec:num_res}, and we conclude our analysis in Sec.~\ref{sec:conclusion}.

    \vspace{-0.5cm}
		\section{Processes}
\label{sec:processes}
\begin{figure}[H]
	\centering
	\includegraphics[width=0.25\linewidth]{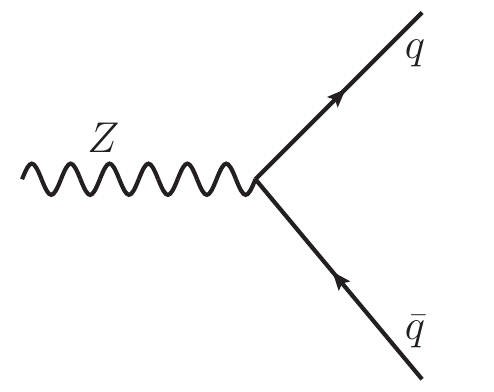}
	\includegraphics[width=0.25\linewidth]{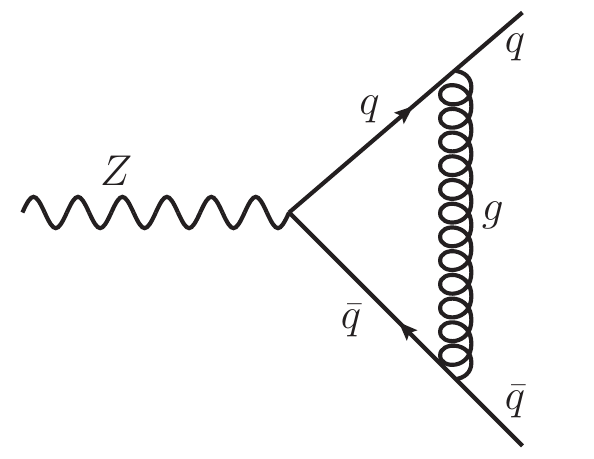}
    \includegraphics[width=0.48\linewidth]{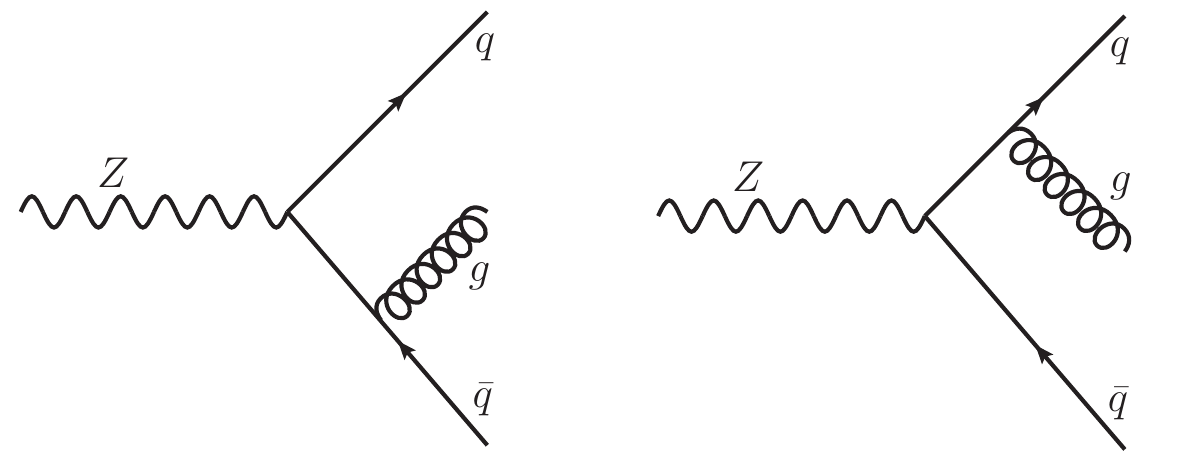}\\
 (a)\hspace{4.5cm}(b)\hspace{4.4cm}(c)\hspace{4cm}(d)
	\caption{(a) Tree-level diagram, (b) virtual correction, and (c), (d) real emission diagram for the process $Z\to q\bar{q}$.}
	\label{fig:zqqct}
\end{figure}

\begin{figure}[H]
			\centering
			\includegraphics[width=0.52\linewidth]{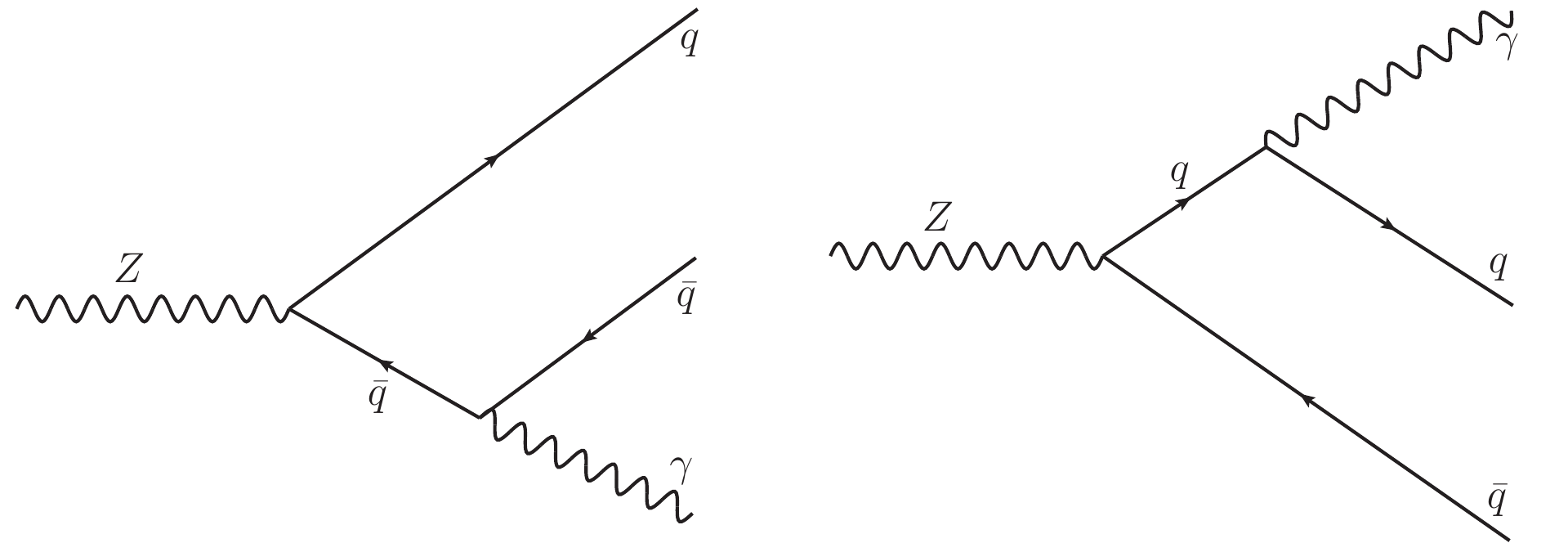}\\
			(a)\hspace{4.0cm}(b)
			\caption{Tree-level diagrams for the process $Z\to q\bar{q}\gamma$.}
 
			\label{fig:tree_2}
		\end{figure}

\begin{figure}[H]
			\centering
			\includegraphics[width=0.7\linewidth]{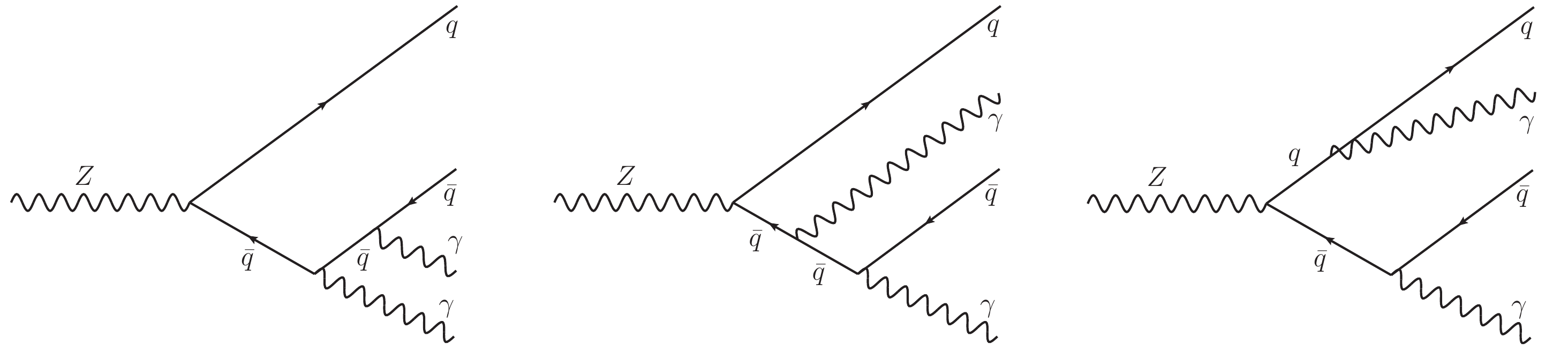}\\
           (a)\hspace{3.8cm} (b)\hspace{3.8cm} (c)
             \includegraphics[width=0.7\linewidth]{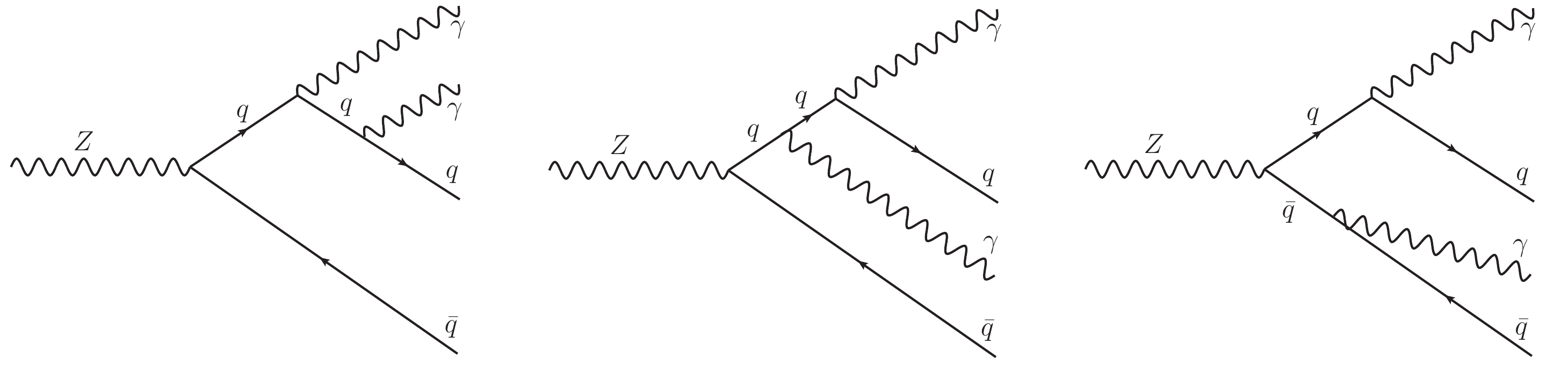}\\
             (d)\hspace{3.8cm} (e)\hspace{3.8cm} (f)
			\caption{Tree-level diagrams for the process $Z\to q\bar{q}\gamma\gamma$.}
			\label{fig:tree_3}
		\end{figure}

  We are interested in studying the processes: (a) $Z\rightarrow q \bar{q}$, (b) $Z\rightarrow q \bar{q} \gamma$, and (c) $Z\rightarrow q \bar{q} \gamma \gamma$. 
  We compute Z boson's inclusive NLO QCD decay widths for these decay channels. 
  The tree-level diagrams for these processes are displayed in Fig.~\ref{fig:zqqct}(a),~\ref{fig:tree_2}, and \ref{fig:tree_3}. 
  There is only one tree-level diagram for the process $Z\rightarrow q \bar{q}$ (Fig.~\ref{fig:zqqct}(a)).  
  Since $\gamma$ can be emitted from either fermion or anti-fermion leg, 
  there are two and six tree-level diagrams for the processes $Z\rightarrow q \bar{q} \gamma$ (Fig.~\ref{fig:tree_2}) and $Z\rightarrow q \bar{q} \gamma \gamma$  (Fig.~\ref{fig:tree_3}), respectively. 

\begin{figure}[H]
			\centering
			\includegraphics[width=1.05\linewidth]{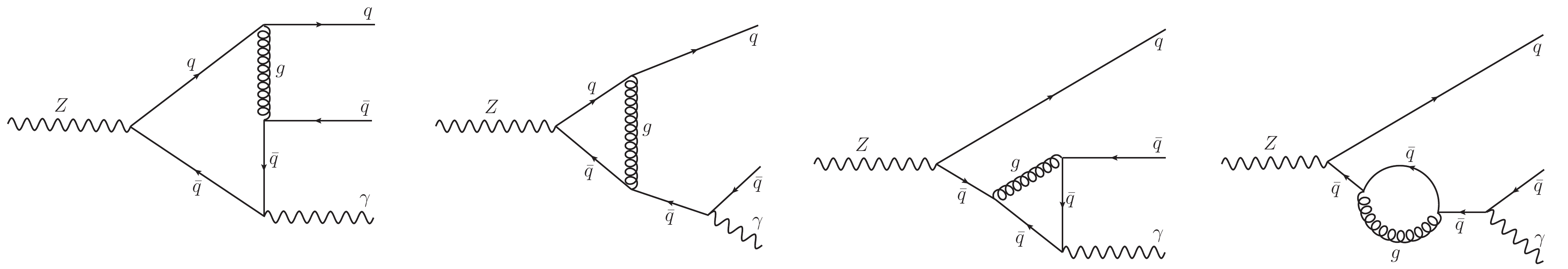}\\
			\hspace{1.5cm}(a)\hspace{4.0cm}(b)\hspace{4.0cm}(c)\hspace{3.7cm}(d)
			\caption{One-loop diagrams for the process $Z\to q\bar{q}\gamma$ corresponding to the tree-level diagram in Fig.~\ref{fig:tree_2}(a). There are similar diagrams (not shown) corresponding to the Fig.~\ref{fig:tree_2}(b).}
			\label{fig:loop_2}
		\end{figure}

\begin{figure}[H]
			\centering
			\includegraphics[width=1.05\linewidth]{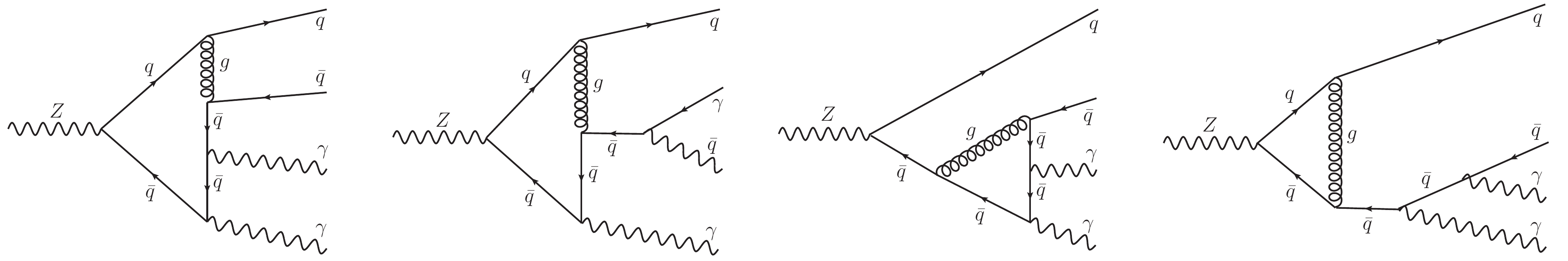}\\
           (a)\hspace{3.8cm} (b)\hspace{3.8cm} (c)\hspace{3.8cm}(d)
             \includegraphics[width=1.05\linewidth]{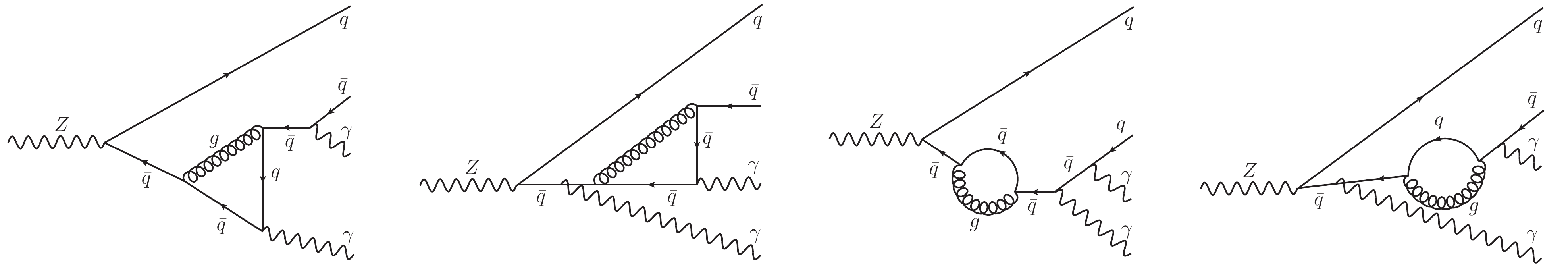}\\
             (e)\hspace{3.8cm} (f)\hspace{3.8cm} (g)\hspace{3.8cm}(h)
			\caption{one-loop diagrams for the process $Z\to q\bar{q}\gamma\gamma$ corresponding to Fig.~\ref{fig:tree_3}(a). 
			There are similar diagrams (not shown) corresponding to the Fig.~\ref{fig:tree_3}(b)-\ref{fig:tree_3}(f).}
			\label{fig:virtualqqaa}
		\end{figure}

  To compute the NLO corrections, one has to compute virtual, counterterm (CT), and real emission diagrams. 
  There are a few virtual diagrams associated with these three processes. 
  For the process $Z\rightarrow q \bar{q}$, as the kinematics is $1\rightarrow 2$, we have triangle and bubble-type diagrams. 
  Similarly, for the process $Z\rightarrow q \bar{q}\gamma$, as the kinematics is $1\rightarrow 3$, there are box, triangle,  and bubble-type diagrams; 
  and for the process $Z\rightarrow q \bar{q}\gamma\gamma$, as the kinematics is $1\rightarrow 4$, there are pentagon, box, triangle,  and bubble-type diagrams.
  There are total $1$, $8$, and $48$ virtual diagrams for $Z\rightarrow q \bar{q}$, $Z\rightarrow q \bar{q}\gamma$, and $Z\rightarrow q \bar{q}\gamma\gamma$ processes, respectively. A few generic virtual diagrams have been shown in Fig.~\ref{fig:zqqct}(b), Fig.~\ref{fig:loop_2}, and Fig.~\ref{fig:virtualqqaa} for these processes.

\begin{figure}[H]
			\centering
			\includegraphics[width=0.7\linewidth]{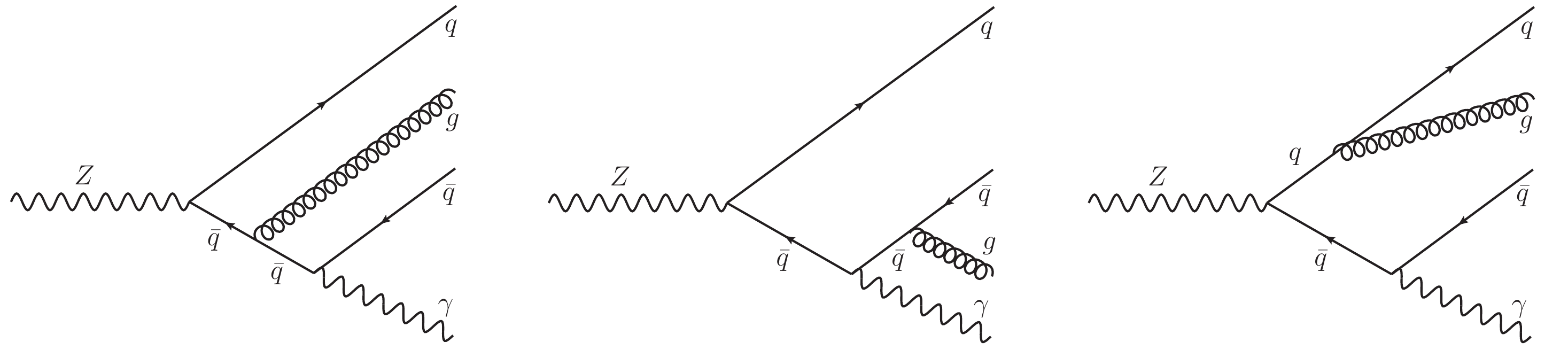}\\
            (a)\hspace{4cm}(b)\hspace{4cm}(c)
			\caption{Real emission diagrams corresponding to the Fig.~\ref{fig:tree_2}(a). There are similar diagrams (not shown) corresponding to the Fig.~\ref{fig:tree_2}(b).}
			\label{fig:rem_2}
		\end{figure}

\begin{figure}[H]
			\centering
			\includegraphics[width=1.05\linewidth]{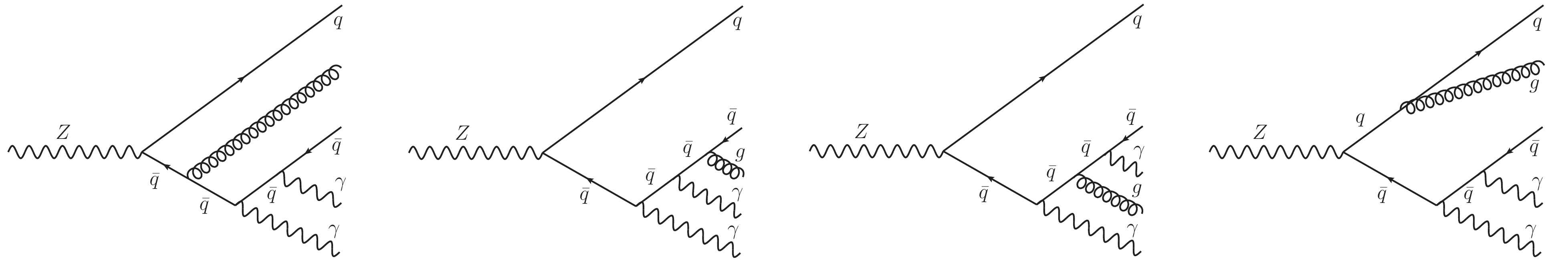}\\
			\hspace{1.0cm}(a)\hspace{4.2cm}(b)\hspace{4.2cm}(c)\hspace{3.8cm}(d)
			
			\caption{Real emission diagramns corresponding to the Fig.~\ref{fig:tree_3}(a). There are similar diagrams (not shown) corresponding to Fig.~\ref{fig:tree_3}(b)-\ref{fig:tree_3}(f).}
			\label{fig:rem_3}
		\end{figure}

For an inclusive observable, one needs to calculate the contribution of radiation diagrams along with the virtual contribution.  A gluon can be radiated from external and internal quarks (anti-quarks).  The real emission processes for these three cases are, respectively, $Z\rightarrow q \bar{q} g$, $Z\rightarrow q \bar{q} \gamma g$, and $Z\rightarrow q \bar{q} \gamma \gamma g$.  There are only $2$ tree-level diagrams for the process $Z\rightarrow q \bar{q} g$ as shown in the Fig.~\ref{fig:zqqct}c and \ref{fig:zqqct}d. The process $Z\rightarrow q \bar{q} \gamma g$, involves $6$ tree-level diagrams, some of which are shown in Fig.~\ref{fig:rem_2}. Similarly, there are a total of $24$ tree-level diagrams for the process $Z\rightarrow q \bar{q} \gamma \gamma g$. The gluon emission diagrams associated with one tree-level diagram of the process $Z\rightarrow q \bar{q} \gamma \gamma$ have been shown in Fig.~\ref{fig:rem_3}. One also needs to compute the CT diagrams along with the virtual and real emission diagrams.  We discuss the renormalization and CT diagrams in Sec.~\ref{subsec:uv_ren}.

\section{Computation}
\label{sec:computation}
For all three processes ((a), (b), and (c)) together, there are a total of $9$ tree-level, $57$ virtual, and $32$ real emission diagrams.  We have generated these diagrams using the {\tt Mathematica} package {\tt FeynArts}~\cite{Hahn:2000kx, Kublbeck:1990xc}.
We have computed amplitudes for tree-level as well as loop-level diagrams using the spinor helicity techniques~\cite{Peskin:2011in}. We have taken external quarks and anti-quarks as massless. One can compute helicity amplitudes for all
helicity configurations of quarks and photons. 
The tree-level amplitudes can be computed in terms of the Lorentz invariant spinor product $\langle pq\rangle$ or $[pq]$~\cite{Peskin:2011in, Kleiss:1986qc}, where $p$, $q$ are the momenta of the spinors and square-triangle brackets are the specific helicity states of the spinors. Along with the spinor product, the vector current of the form $\langle p\gamma^\mu q]$ (or $[p\gamma^\mu q\rangle$) is needed for virtual amplitude computation.
We have obtained the functional form of the $\langle pq\rangle$ and $\langle p\gamma^\mu q]$~\cite{Das:2023gsh}. We have verified that these functional forms satisfy the helicity identities.

We use {\tt FeynArts} and {\tt FormCalc}~\cite{Hahn:1998yk} packages to get raw Feynman amplitudes. We calculate helicity amplitudes from these raw amplitudes using our in-house {\tt FORM}~\cite{Vermaseren:2000nd} routine. This routine converts raw amplitudes to scalar objects which can be computed numerically.
The {\tt FORM} routine writes amplitudes in terms of scalar and tensor integral functions, spinor products, dot products among vector currents, polarizations, and momenta. At the loop level, there are UV and IR divergences. The radiation diagrams
also have IR divergences. For these amplitude computations, we use  
 the 't Hooft-Veltman dimensional scheme, known as the HV scheme. In this scheme, the part associated with the loop has been computed in the $d$-dimension, and the rest part has been computed in the  $4$-dimension. 
We compute tensor integrals using our in-house tensor reduction routines {\tt OVReduce}~\cite{Agrawal:2012df, Agrawal:1998ch}. This routine has been written using Oldenborgh-Vermaseren basis~\cite{vanOldenborgh:1989wn}. For the scalar integrals, we use {\tt OneLoop}~\cite{vanHameren:2010cp} package.

 Finally, to perform the phase space integration, we use the package advanced Monte-Carlo integration {\tt AMCI}~\cite{Veseli:1997hr}.  This package implements  {\tt VEGAS}~\cite{Lepage:1977sw} algorithm using the parallel virtual machine ({\tt PVM})~\cite{10.7551/mitpress/5712.001.0001} for {\tt AMCI} to compute the integrand parallelly across the nodes. To generate a phase-space configuration, we use {\tt RAMBO}~\cite{Kleiss:1985gy} routine. The computation is relatively more tedious for the processes $Z\rightarrow q q\gamma\gamma$ due to two photons in the final state. Apart from
four particles in the final state, the process has a large number of possible helicity configurations and three-tensor
pentagon-type diagrams.
 It takes a fair amount of CPU time with a large number of phase space points to give stable results. We also face the usual numerical instabilities with the pentagon amplitudes. We see that there are points in phase space where the ratios of virtual correction to leading order are unnaturally large. These kinds of unwanted points can destroy the convergence of the integrals. As usual, the trick is to throw away these very few phase space points where the contribution is huge~\cite{Greiner:2010dth}. We have used a local $K$-factor for this purpose, as given below
 \begin{align}
     K=\Bigg|\frac{2\mathcal{M}^*_{\text{virt}}\mathcal{M}_{\text{LO}}}{|\mathcal{M}_{\text{LO}}|^2}\Bigg|.
 \end{align}
 We see a plateau region for the final results between $4<K<6$, so we set $K=5$ for our computations. We do not see these issues for other virtual amplitudes other than pentagon amplitudes.

\subsection {UV renormalization}
\label{subsec:uv_ren}
There are triangle-type virtual diagrams for all three processes.  There are many triangle diagrams, such as the one in Fig.~\ref{fig:zqqct}(b) for $Z\to q\bar{q}$; Fig.~\ref{fig:loop_2}(b) and \ref{fig:loop_2}(c) for the process $Z\to q\bar{q}\gamma$; Fig.~\ref{fig:virtualqqaa}(d), \ref{fig:virtualqqaa}(e), and \ref{fig:virtualqqaa}(f) for the process $Z\to q\bar{q}\gamma\gamma$, etc., where two fermion lines are inside the loop. The tensor integrals of these diagrams are UV divergent.  The bubble diagrams in Fig~\ref{fig:loop_2}(d), \ref{fig:virtualqqaa}(g), \ref{fig:virtualqqaa}(h) are also UV divergent as the bubble-type integrals are UV divergent. One can introduce the CT diagrams systematically at each vertex and fermionic propagator in tree-level diagrams to remove the UV poles from the amplitudes. 
As we know, one-loop QCD renormalization does not renormalize the electric charge. However, wavefunction (w.f.) and mass renormalization can occur. In our calculation, we have adopted the on-shell renormalization scheme. Here, the fermions are treated as massless, and the w.f. renormalization vanishes for massless fermions.

\subsection{IR Singularities and Dipole Subtraction}
\label{subsec:ir_ds}
In this process, both virtual and real emission diagrams are separately IR singular at $4$-dimension, but their sum is finite. The IR singularity appears as $\frac{1}{\epsilon}$ in dimensional regularization and as $\log(m/Q)$ in mass regularization, where $\epsilon=4-d$ and $m$ is the parton mass, with $Q$ representing a large scale.
Performing phase-space integrals analytically for multiparticle processes at colliders is tedious and often nearly impossible. Therefore, we perform numerical Monte-Carlo integration to obtain stable results.
The virtual and real contributions have to be integrated separately as they have a different number of final states. As there is a mismatch of the final state phase-space integral, the advantage of pole cancellation can not be taken for an exclusive observable. The IR singularities that appear in the intermediate steps should be taken care of separately, and then the numerical integration can be performed safely. Hence, one subtraction scheme is desired to perform the phase-space integration numerically. 

We adopt the Catani-Saymour (CS) dipole subtraction scheme~\cite{Catani:1996vz} to handle the IR singularities from virtual and real emission amplitudes. The full  decay width can be written as 
\begin{equation}
    \Gamma^{\rm NLO}=\int d\Gamma^{\rm NLO}=\int_{m+1} d\Gamma^R+\int_m d\Gamma^V.
    \label{eq:gammanlo}
\end{equation}

Here, $\Gamma^R$ is the real emission contribution, which has been computed with an extra particle in the final state phase-space, and $\Gamma^V$ is the virtual contribution.  The two integrals in Eq.~\eqref{eq:gammanlo} are separately divergent, but their sum is finite in $4$-dimension.  According to the CS scheme, a \textit{local} counterterm is added to the virtual part, and the same term is subtracted from the real part. This local counterterm has the same \textit{pointwise} singular behavior as $d\Gamma^R$ itself. This can be expressed as
\begin{equation}
    \Gamma^{\rm NLO}=\int_{m+1}\big[d\Gamma^R-d\Gamma^A\big]+\int_m \big[d\Gamma^V+\int_1 d\Gamma^A\big].
    \label{equ:dp_ct}
\end{equation}
One can perform one-particle phase space integral over $d\Gamma^A$ in $d$-dimension and absorb the IR poles from $d\Gamma^V$. Once we remove the IR poles from $d\Gamma^V$ and as $d\Gamma^A$ has the same pointwise singular behavior as $d\Gamma^R$, we take limit $\epsilon\rightarrow 0$ for both integrals in Eq.~\eqref{equ:dp_ct} and perform the integrals in $4$-dimension. The term $d\Gamma^A$ will give the dipole terms $dV_{\text{dipole}}$, summed over all dipoles. The term $\int_1 d\Gamma^A$ will give an integrated dipole term, referred to as the $\mathbf{I}$-term. Eq.~\eqref{equ:dp_ct} can therefore be written as 
\begin{equation}
  \Gamma^{\rm NLO}=\int_{m+1}\big[d\Gamma^R-\sum_{\text{dipoles}}d\Gamma^B\otimes dV_{\text{dipole}}\big]_{\epsilon=0}+\int_m \big[d\Gamma^V+d\Gamma^B\otimes \mathbf{I}\big]_{\epsilon=0},
  \label{equ:nlo_gamma}
\end{equation}
where $\Gamma^B$ is the Born-level decay width.
One has to calculate the dipole terms $dV_{\text{dipoles}}$ and  $\mathbf{I}$-term to get IR-safe amplitude separately for $m$ and $m+1$ partonic phase-space.

Following Ref.~\cite{Catani:1996vz}, the insertion kernel is given by
\begin{align}
     {\bf{I}}(\{p_f\};\epsilon)=-\frac{\alpha_s}{2\pi}\frac{1}{\Gamma(1-\epsilon)}\sum_I\frac{1}{{\bf{T}}_I^2}\mathcal{V}_I(\epsilon)\sum_{J\neq I}{\bf{T}}_I.{\bf{T}}_J\Bigg(\frac{4\pi\mu^2}{2p_J.p_I}\Bigg)^\epsilon,
     \label{equ:I_exp}
 \end{align}
 where the indices $I$, $J$ run over the final state partons (for our processes), ${\bf{T}}_I$ is the color charge for $I$-th parton and 
 \begin{align}
     \mathcal{V}_I(\epsilon)={\bf{T}}_I^2\Bigg(\frac{1}{\epsilon^2}-\frac{\pi^2}{3}\Bigg)+\gamma_I\frac{1}{\epsilon}+\gamma_I+K_I+\mathcal{O}(\epsilon),
     \label{equ:Nu_eps}
 \end{align}
 where the factors $K_I$, $\gamma_I$ are given in Ref.~\cite{Catani:1996vz}.
 
 The relevant partons in our processes for integrated dipole term {\bf {I}} are final state quarks (anti-quarks). For these processes, the {\bf {I}}-term can be written as
    \begin{align}
        \mathbf{I}(\{p\};\epsilon)=\frac{\alpha_s}{2\pi}\frac{(4\pi)^\epsilon}{\Gamma(1-\epsilon)}(-2C_F)&\Big[\frac{1}{\epsilon^2}+\frac{1}{\epsilon}\Big\{\frac{3}{2}+\log\frac{\mu^2}{2p_q.p_{\bar{q}}}\Big\}\nonumber\\
        &+\Big\{-\frac{\pi^2}{2}+5+\frac{3}{2}\log\frac{\mu^2}{2p_q.p_{\Bar{q}}}+\frac{1}{2}\Big(\log\frac{\mu^2}{2p_q.p_{\Bar{q}}}\Big)^2+\mathcal{O}(\epsilon)\Big\}\Big].
 \label{equ:I_term}
    \end{align}
This {\bf {I}}-term will remove all IR singularities from the virtual amplitude and add a finite contribution. The {\bf {I}}-term is convoluted with the Born-level amplitude square with $m$-phase-space final states.

We have only two partons (quarks) in these processes in the final state. Following the CS scheme, we calculate a total of two final-final (FF) dipole terms as there are no initial partons for these processes. We write the local counterterm as the sum over FF dipoles,
   \begin{align}
        \int_{m+1}d\Gamma^A=\sum_{k\neq ij}\:\int d\Pi_m\: \mathcal{D}_{ij,k}^{FF} \:J_m,
    \label{equ:dipole_ct}
    \end{align}
    
    where $d\Pi_m$ is the $m$-particle Lorentz invariant phase-space and $J_m$ is the infrared safe jet function which means  $J_{m+1}\rightarrow J_m$ when $m+1$ and $m$-partonic configuration are kinematically degenerate. The FF dipole is given by 
    \begin{align}
        \mathcal{D}_{ij,k}^{FF}=-\frac{1}{2p_i.p_j}\langle ...,\Tilde{ij},...,\Tilde{k},...|\frac{{\bf{T}}_k.{\bf{T}}_{ij}}{{\bf{T}}^2_{ij}} {\bf{V}}_{ij,k}|...,\Tilde{ij},...,\Tilde{k},... \rangle,
    \end{align}
where parton $\Tilde{ij}$ is called the emitter and parton $\Tilde{k}$ is called the spectator. ${\bf{T}}_{ij}$ and ${\bf{T}}_j$ represent the color charges of the emitter and spectator. The ${\bf{V}}_{ij,k}$ is the splitting matrix which depends on the kinematical variables $y_{ij,k}$, $\Tilde{z}_i$, $\Tilde{z}_j$which are given as 
\begin{align}
    y_{ij,k}=\frac{p_i.p_j}{p_i.p_j+p_j.p_k+p_k.p_i},\quad \Tilde{z}_{j}=\frac{p_j.p_k}{p_j.p_k+p_i.p_k}=1-\Tilde{z}_{i}.
\end{align}
    For our processes, the spitting matrix in the emitter helicity basis ($s$ and $s^\prime$) is
    \begin{align}
        \langle s|{\bf{V}}_{q_ig_j,k}|s^\prime \rangle=8\pi\mu^{2\epsilon}\alpha_S\: C_F\Big[ \frac{2}{1-\Tilde{z}_i(1-y_{ij,k})}-(1+\Tilde{z}_i)-\epsilon(1-\Tilde{z}_i)\Big]\delta_{ss^\prime}.
    \end{align}
The dipole terms are calculated with $m$-partonic phase space derived from real $(m+1)$-partonic phase space. We compute the reduced mapped momenta for dipole terms following the CS scheme.

As virtual corrections are not always positive definite, depending on the subtraction term, it might be possible that either of the two terms ($\Gamma^R-\Gamma^{\rm dipoles}$ and $\Gamma^V+\Gamma^{\rm I}$) in Eq.~\eqref{equ:nlo_gamma} will not be positive definite. Also, integrating dipole terms along with real emissions diagrams is quite CPU time-consuming.
To address these issues, we introduce a parameter called $\alpha\in (0,1]$ in our dipole computation which will restrict the phase space of dipole terms in our dipole computation. Inserting $\alpha$ in the calculation is for numerical advantage and saves CPU time. $\alpha=1$ means no restriction on phase space i.e., full phase space integral for dipoles.
One can check the consistency of the results (total NLO) by varying $\alpha$. Introducing $\alpha$ in dipoles, the Eq.~\ref{equ:dipole_ct} becomes
\begin{align}
        \int_{m+1}d\Gamma^A=\sum_{k\neq ij}\:\int d\Pi_m\: \mathcal{D}_{ij,k}\:\Theta(y_{ij,k}< \alpha) \:J_m,
   \label{equ:dipole_ct_alp}
    \end{align}
 with $\alpha\in(0,1]$. 

 The introduction of $\alpha$ phase space cut will also redefine the kernel ${\bf{I}}$ and following Ref[] we replace $\mathcal{V}_i(\epsilon)\rightarrow \mathcal{V}_i(\alpha,\epsilon)$ in Eq.~\ref{equ:Nu_eps}, with
 
 \begin{align}
     \mathcal{V}_i(\alpha,\epsilon)={\bf{T}}_i^2\Bigg(\frac{1}{\epsilon^2}-\frac{\pi^2}{3}\Bigg)+\gamma_i\frac{1}{\epsilon}+\gamma_i+K_i(\alpha)+\mathcal{O}(\epsilon),
 \end{align}
 where
 \begin{align}
     K_i(\alpha)=K_i-{\bf{T}}^2_i \:\text{ln}^2\alpha + \gamma_i\:(\alpha-1-\text{ln}\alpha).
 \end{align}
 
 We choose $\alpha=0.5$ consistently in our computation to save the CPU time. 
\subsection{Checks}
\label{subsec:checks}
We made several checks to validate our computations. These checks are listed below.

\begin{enumerate}
    \item {\it {Pole cancellation}}: With the {\bf {I}}-term given in the Eq.~\ref{equ:I_term}, we have checked that it removed all poles from virtual amplitudes for each process.
    \item {\it{Cancellation with dipoles}}: The dipole terms display explicitly the same singular behavior as the real emission amplitude in the collinear and soft regions in each process.  Two numerical numbers corresponding to real and dipole terms are large in the singular region, but their subtraction lies in the regular region. This leads to stable dipole subtracted real emission contribution. This has been the case in each process.
    \item {\it $\alpha$ parameter}: Introduction of $\alpha$ parameter in dipole subtraction not only saves CPU time but also helps to validate the implementation of dipole subtraction. By varying $\alpha\in (0,1]$, we see consistent results for all processes within statistical uncertainties. 
    \item {\it {Gauge invariance}}: There are massless gauge bosons (photons) in the process (2) and (3) at the LO. For all real emission processes, in addition to the photons, there are also gluons in the final state. We have checked the Ward-Identity by replacing the gauge boson polarization with its momentum. This gives a vanishing contribution. This check has been done with the LO born-level, the NLO virtual, and the NLO real emission contributions for each process except the born-level of $Z\rightarrow q \bar{q}$ process where there is no massless gauge boson in the external state.
\end{enumerate}

Additionally, we have checked that the born-level decay widths have an agreement with {\tt MadGraph}. We have also verified that the NLO corrected decay width for the process $Z\rightarrow q \bar{q}$ exactly matches with the Ref.~\cite{Beenakker:1988pv}.

\section{Numerical Results}
\label{sec:num_res}
We calculate the total decay widths for the processes $Z\rightarrow q{\bar q}$,  \:$Z\rightarrow q{\bar q}\gamma$, \:and \:$Z\rightarrow q{\bar q}\gamma\gamma$. There are five sub-processes corresponding to five massless quarks (we take $b$-quark as massless) for each process.
The total decay width is computed by summing twice the partial width for up-type quark channels and three times the partial width for down-type quark channels.

In order to evaluate the partial decay widths of the said processes, the model parameters in the EW and strong sectors must be specified. These include the coupling constants $\alpha_s(M_Z^2)$, $\alpha(M_Z^2)$, the weak scale, and the $Z$ boson line shape parameter, or alternatively, the $Z$ boson decay width $\Gamma_Z$~\cite{Ge:2024pfn, Dawson:2019clf, ParticleDataGroup:2018ovx, Novikov:1999af},
\begin{align}
    G_F =& 1.1663\times 10^{-5}\ \text{GeV}^{-2},\hspace{1.8cm} \alpha(M_Z^2) =1/ 128.946,\nonumber\\
    M_Z =& 91.1876\ \text{GeV}, \hspace{3.2cm} \alpha_s(M_Z^2) = 0.1181,\nonumber\\
   M_W=& 80.379\ \text{GeV},\hspace{4.25cm} \Gamma_Z = 2.4952\ \text{GeV}.
\end{align}

In addition to $\Gamma_Z$, the other three parameters can be derived from the fine$-$structure constant $\alpha$, the Fermi constant $G_F$, and the $Z$ boson mass $M_Z$. These ($\alpha$, $G_F$, $M_Z$) are the most precisely measured EW parameters. Recently, various EW input parameter sets for SM Effective Field Theory (SMEFT) predictions at the LHC have been discussed~\cite{Brivio:2021yjb}.

We do not put any kinematical cuts for the process $Z\rightarrow q\bar{q} $, but we put below kinematical cuts on photons for the processes $Z\rightarrow q \bar{q}\gamma$ and $Z\rightarrow q \bar{q}\gamma\gamma$ for total decay width
\begin{equation}
    p_T=10\:{\text {GeV and}}  \:\:\Delta R=0.4.
    \label{equ:ct_photon}
\end{equation}
The $\Delta R$ separation cuts are between photon and final state partons in ($\eta-\phi$) plane. These cuts are necessary to have finite decay widths. The total widths for the processes $Z\rightarrow q{\bar q}$,  \:$Z\rightarrow q{\bar q}\gamma$, \:and \:$Z\rightarrow q{\bar q}\gamma\gamma$ have been listed in Tab.~\ref{tab:pw_tot}, where $\Gamma^{\text{LO}}$ and $\Gamma^{\text{NLO}}$ represent the LO and NLO decay widths, respectively. We define the relative increment as ${\textbf {RI}}=\frac{\Gamma^{\text{NLO}}-\Gamma^{\text{LO}}}{\Gamma^{\text{LO}}}\times 100\%$. As mentioned in Tab.~\ref{tab:pw_tot}, the decay widths for $Z\rightarrow q{\bar q}$,  \:$Z\rightarrow q{\bar q}\gamma$, \:and \:$Z\rightarrow q{\bar q}\gamma\gamma$ at LO are $1757.36$~MeV, $6.30$~MeV, and $11.30$~keV, respectively. At NLO, the decay widths are $1823.61$~MeV, $5.92$~MeV, and $9.90$~keV for the prosesses $Z\rightarrow q{\bar q}$,  \:$Z\rightarrow q{\bar q}\gamma$, \:and \:$Z\rightarrow q{\bar q}\gamma\gamma$, respectively. For the process $Z\rightarrow q{\bar q}$, the NLO correction is positive, with \textbf{RI}$=$ $3.77\%$. In contrast, for the process $Z\rightarrow q{\bar q}\gamma$, \:and \:$Z\rightarrow q{\bar q}\gamma\gamma$, the NLO corrections are negative, with \textbf{RI} values of $-6.03\%$ and $-12.39\%$, respectively.

\begin{table}[H]
\centering
\begin{tabular}{|c|c|c|c|}

  \hline
  \multirow{2}{*}{\textbf{Process}} & \multicolumn{2}{|c|}{\textbf{Partial decay width}} & \multirow{2}{*}{\textbf{RI ($\%$)}} \\
  \cline{2-3}
  & $\Gamma^{\rm LO}$ (MeV) & $\Gamma^{\rm NLO}$ (MeV) & \\
  \hline
  $Z\to q\bar{q}$ & 1757.36 & 1823.61 & 3.77 \\
  \hline
  \hline
  $Z\to q\bar{q}\gamma$ & 6.30 & 5.92 & $-6.03$ \\
  \hline
  \hline
  $Z\to q\bar{q}\gamma\gamma$ &  $1.13\times 10^{-2}$  &  $9.90\times 10^{-3}$  & $-12.39$  \\
  \hline
\end{tabular}
\caption{Partial decay width and the relative increment of the NLO corrections with respect to the Born-level for the processes $Z\to q\bar{q}$, $Z\to q\bar{q}\gamma$, and $Z\to q\bar{q}\gamma\gamma$.}
    \label{tab:pw_tot}
\end{table}

The jet-level decay widths corresponding to the processes $Z\rightarrow q{\bar q}$,  \:$Z\rightarrow q{\bar q}\gamma$, \:and \:$Z\rightarrow q{\bar q}\gamma\gamma$ are presented in Tab.~\ref{tab:pw_jet}. Note that the LO decay width for $Z\to$ 1-jet is not possible. Similarly, the LO decay widths for $Z\to $3-jet, 3-jet+$\gamma$, and 3-jet+$2\gamma$ are not possible. However, at NLO, all the jet-level decay widths are possible.

\begin{table}[H]
\centering
\begin{tabular}{|c|c|c|c|}

  \hline
  \multirow{2}{*}{\textbf{Process}} & \multicolumn{2}{|c|}{\textbf{Partial decay width}}& \multirow{2}{*}{\textbf{RI ($\%$)}}\\
  \cline{2-3}
  & $\Gamma^{\rm LO}$ (MeV) & $\Gamma^{\rm NLO}$ (MeV)&    \\
  \hline
  \hline
  $Z\to 1$-Jet & $-$ & 37.14 & $-$ \\
  \hline
  $Z\to 2$-Jet & 1714.69 & 1336.48 & $-22.06$\\
  \hline
  $Z\to 3$-Jet & $-$ & 419.38 & $-$\\
  \hline
  \hline
  $Z\to 1$-Jet+$\gamma$ & 1.60 & 1.38& $-13.75$  \\
  \hline
  $Z\to 2$-Jet+$\gamma$ & 4.69 & 3.74 & $-20.26$ \\
  \hline
  $Z\to 3$-Jet+$\gamma$ & $-$ & 0.67 & $-$ \\
  \hline
  \hline
  $Z\to 1$-Jet+$2\gamma$ & $4.78\times 10^{-3}$ & $3.26\times 10^{-3}$ & $-31.80$ \\
  \hline
  $Z\to 2$-Jet+$2\gamma$ & $5.59\times 10^{-3}$ & $4.39\times 10^{-3}$& $-21.47$  \\
  \hline
  $Z\to 3$-Jet+$2\gamma$ & $-$ & $3.67\times 10^{-4}$ & $-$ \\  
  \hline
\end{tabular}
\caption{The jet-level decay widths, $Z\to$jet(s), jet(s)+$\gamma$, and jet(s)+$2\gamma$ corresponding to the processes $Z\rightarrow q{\bar q}$,  \:$Z\rightarrow q{\bar q}\gamma$, \:and \:$Z\rightarrow q{\bar q}\gamma\gamma$, respectively.}
    \label{tab:pw_jet}
\end{table}

Given the substantial production cross-section for the $Z$ boson at the HL-LHC ($\sqrt{s}=14$TeV, $\mathcal{L}=3\text{ab}^{-1}$), approximately $\sigma_Z \simeq 55.6$~nb at the LO, we can make a rough estimate of the production cross-sections for the $q\bar{q}\gamma$ and $q\bar{q}\gamma\gamma$ processes at the HL-LHC. For the $q\bar{q}\gamma$ channel, we have the cross-section $\sigma_{q\bar{q}\gamma}=\sigma_Z\times \text{BR}(Z\to q\bar{q}\gamma)\simeq 2.37\times 10^{-3}\times 55.6$~nb $\simeq 0.13$~nb. Similarly, for the $q\bar{q}\gamma\gamma$ channel, $\sigma_{q\bar{q}\gamma\gamma}= \sigma_Z\times \text{BR}(Z\to q\bar{q}\gamma\gamma)\simeq 3.97\times 10^{-6}\times 55.6$~nb $\simeq 0.22$~pb. Therefore, it can be interesting to test these decay channels at the HL-LHC.

\vspace{-0.3cm}
\subsection{Kinematical Distributions}

We are interested in studying the various differential kinematical distributions for these three processes, along with the total decay widths. As there are many partons in the final state, one can use a jet clustering algorithm to define final-state jets.
Depending upon the jet clustering, there can be $1$-jet, $2$-jet, and $3$-jet events. For the process $Z\rightarrow q{\bar q}$, there are only jets in the events. For the processes $Z\rightarrow q{\bar q}\gamma$ and $Z\rightarrow q{\bar q}\gamma\gamma$, one and two photons are produced respectively associated with the jets.

We use the {\tt FastJet}~\cite{Cacciari:2011ma, Cacciari:2005hq} routines to form the jets out of the partons. We choose the $k_t$ algorithm~\cite{Catani:1993hr, Ellis:1993tq} in {\tt FastJet} with the separation $R=0.4$. Along with cuts given in Eq.~\eqref{equ:ct_photon}, we put a $10$ GeV minimum transverse momenta cut on jets. We also demand well-separated jets and photons.
We put minimum $\Delta R=0.4$ separation between the jets and photons, and between the photons.
The {\tt FastJet} and these kinematical cuts are also implemented for dipoles for consistent computation.
These implementations are IR-safe, which means all soft and collinear configurations are allowed for emission, and their cancellations with dipoles are not affected. This guarantees the IR safety of the jet algorithm with the chosen kinematical cuts.
As discussed in Sec.~\ref{subsec:ir_ds}, an auxiliary ($d\Gamma^A$) term is added with the virtual amplitude and subtracted from the real amplitude in dipole subtraction. Although this auxiliary term removes IR singularities safely from virtual as well as real emission amplitudes, it can give some finite contributions to both sides, which vanishes in the final result.
One has to ensure that the finite contribution of this auxiliary term should be canceled bin by bin for differential distribution. For this, we fill up the bins separately for the real emission and dipoles.
As these weights are separately divergent in soft and collinear regions, we have implemented very minimal cuts in these regions. The contribution from these cut regions is negligible compared to the total NLO corrected decay widths.

\begin{figure}[H]
			\centering
			\includegraphics[width=0.38\linewidth]{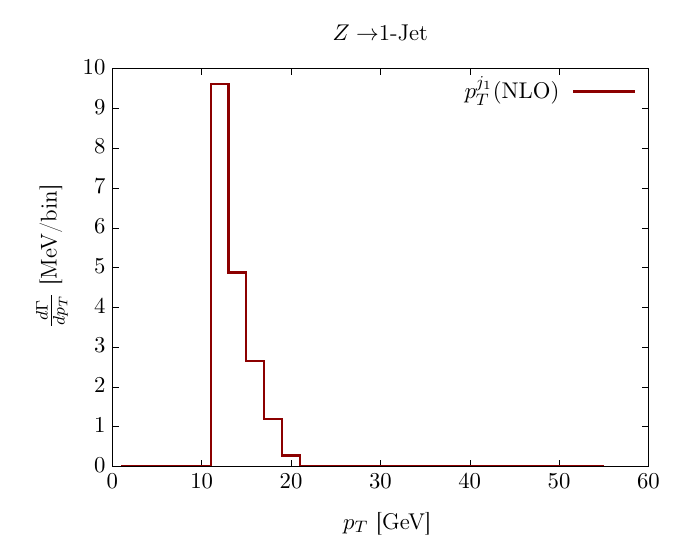}
            \includegraphics[width=0.38\linewidth]{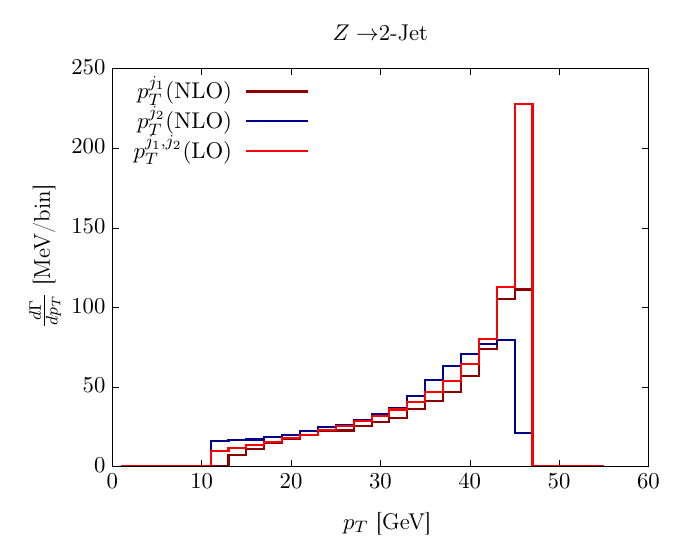}\\
            \hspace{1.0cm}(a)\hspace{5.7cm}(b)\\
            \includegraphics[width=0.38\linewidth]{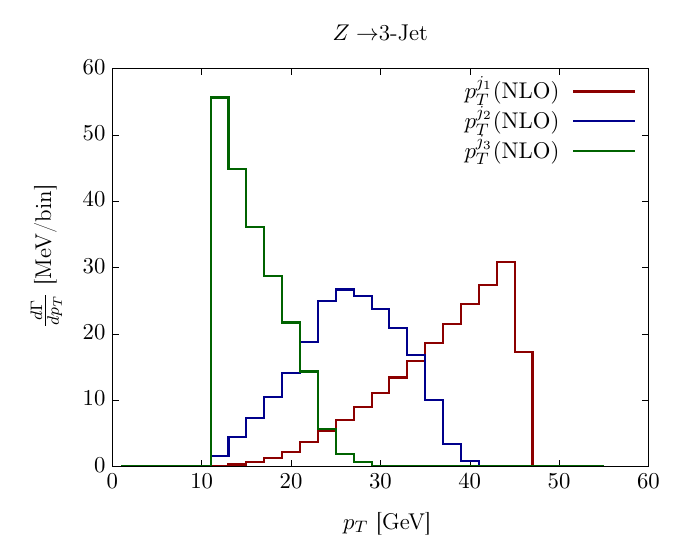}\\
			\hspace{0.8cm} (c) 
	
			\caption{Distributions of the width with respect to $p_T$.}
			\label{fig:qq_pT}
		\end{figure}

\vspace{-0.4cm}

\begin{figure}[H]
			\centering
			\includegraphics[width=0.38\linewidth]{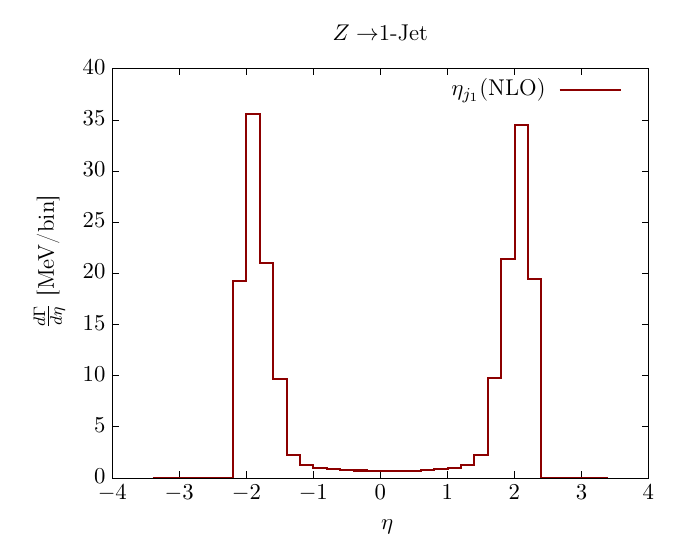}
            \includegraphics[width=0.38\linewidth]{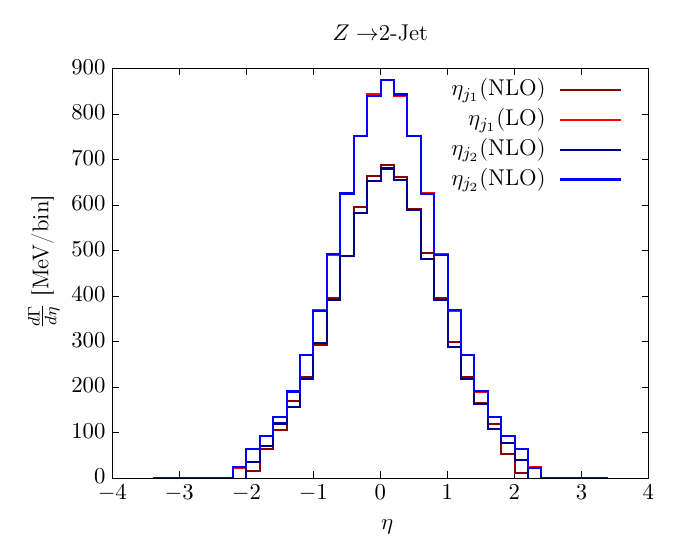}\\
            \hspace{1.0cm}(a)\hspace{5.7cm}(b)\\
            \includegraphics[width=0.38\linewidth]{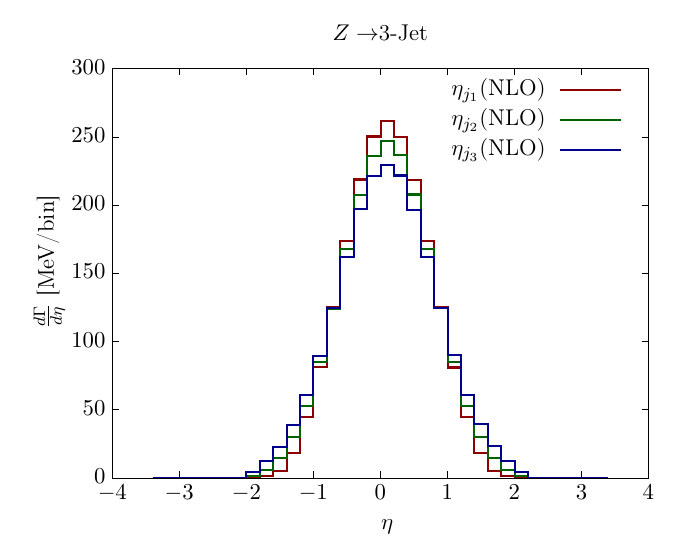}\\
			\hspace{0.8cm}(c) 
	
			\caption{Distributions of the width with respect to $\eta$.}
			\label{fig:qq_Eta}
		\end{figure}

\begin{figure}[H]
			\centering
			\includegraphics[width=0.36\linewidth]{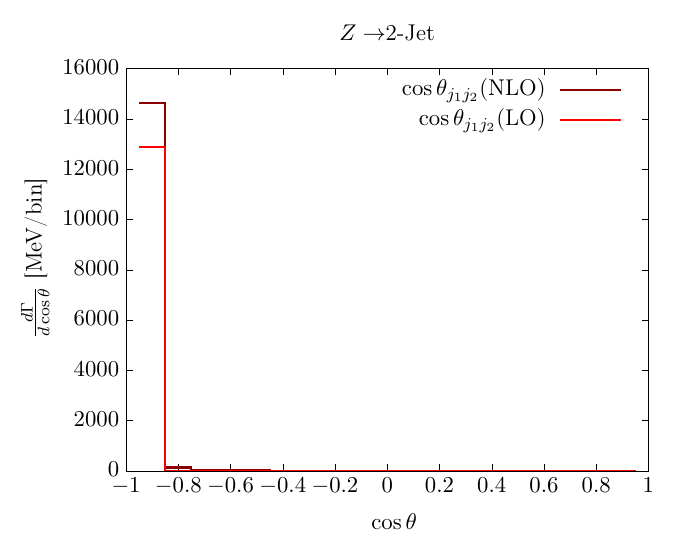}
            \includegraphics[width=0.36\linewidth]{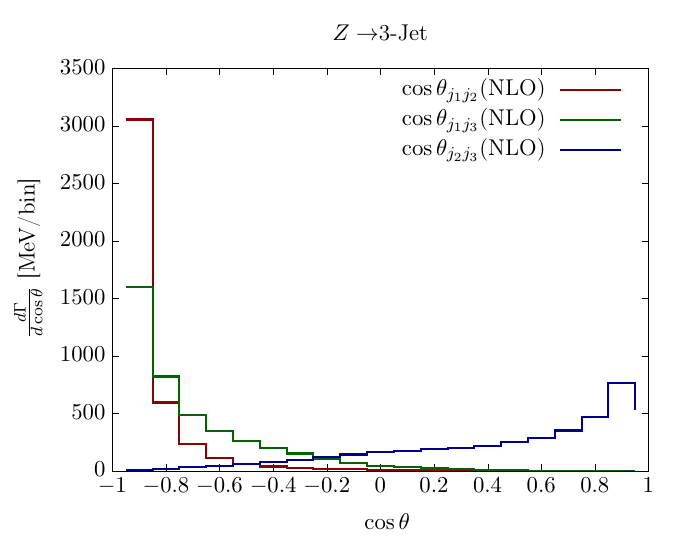}\\
			\hspace{1.0cm}(a)\hspace{5.4cm}(b) 
	
			\caption{Distributions of the width with respect to $\cos\theta$.}
			\label{fig:qq_csT}
		\end{figure}

\begin{figure}[H]
			\centering
			\includegraphics[width=0.36\linewidth]{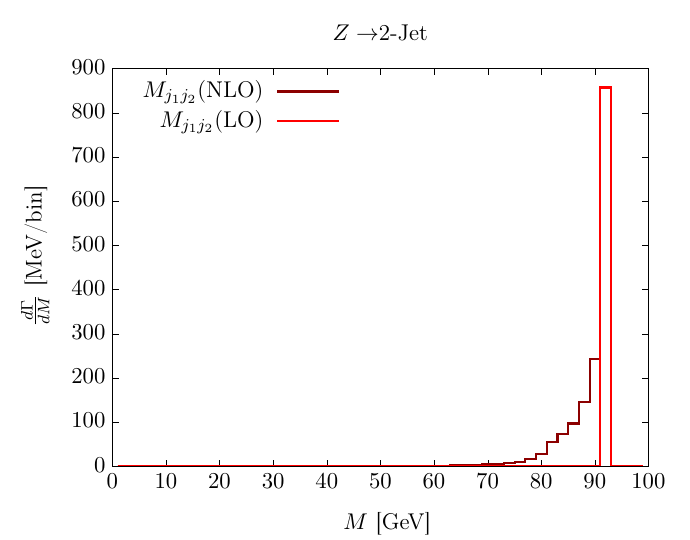}
            \includegraphics[width=0.36\linewidth]{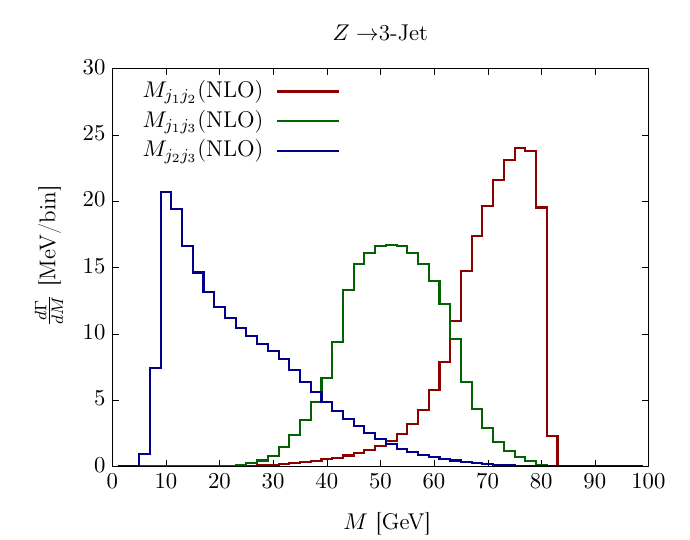}\\
			\hspace{1.0cm}(a)\hspace{5.4cm}(b) 
	
			\caption{Distributions of the width with respect to the mass of a two-jet system.}
			\label{fig:qq_mass}
		\end{figure}

For the process $Z \to q {\bar q}$ at the LO, there will be two jets in the final state. However, at the NLO level, we could have 1, 2, or 3-jet events. In Fig.~\ref{fig:qq_pT}, we have plotted the $p_T$ distributions for each type of event. For the events with multiple jets, for plots, we have ordered jets according to their $p_T$. The width for the 1-jet event is very small and peaked towards the minimum $p_T$ value. For the 2-jet case, as one would expect, the $p_T$ peaked towards $M_Z/2$ for both the LO and NLO cases. In the NLO case, the overall width for the 2-jet case is lower due to the emission of a gluon that leads to one and 3-jet events also. 
The number of 3-jet events is not insignificant. 
As expected, one of the jets, due to a quark or anti-quark, has $p_T$ towards the $M_Z/2$. The jet, due to gluon, has a peak towards a small $p_T$ value. In Fig.~\ref{fig:qq_Eta}, the pseudo-rapidity is plotted. For the 1-jet
case, the jet has a larger pseudo-rapidity; for the 2 and 3-jet cases, the peak is at
zero pseudo-rapidity with a smaller spread for the 3-jet case. In Fig.~\ref{fig:qq_csT},
the cosine of the angle between jets is plotted. As expected, for the 2-jet case, the two jets are
largely back-to-back. For the 3-jet case, the two jets are also back-to-back, except for
the case when gluon is emitted from a (anti)-quark. In Fig.~\ref{fig:qq_mass}, the masses of
two-jet systems are plotted. These distributions behave in the same way as the $p_T$ distributions.

\vspace{-0.3cm}

  \vspace{-1.0cm}

\vspace{-0.25cm}

\begin{figure}[H]
			\centering
			\includegraphics[width=0.36\linewidth]{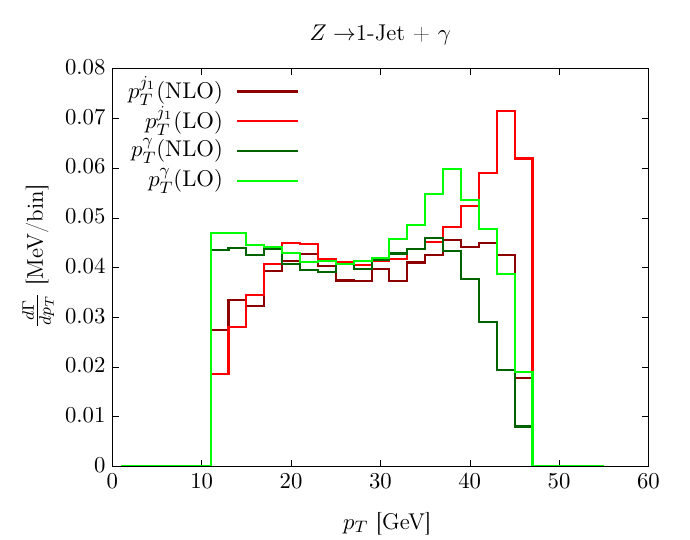}
            \includegraphics[width=0.36\linewidth]{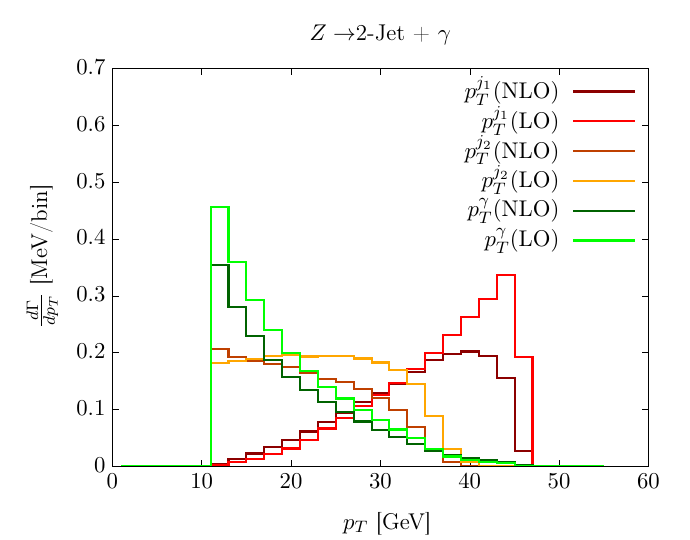}\\
            
			\hspace{1.0cm}(a)\hspace{5.5cm}(b)\\ 
	        \includegraphics[width=0.36\linewidth]{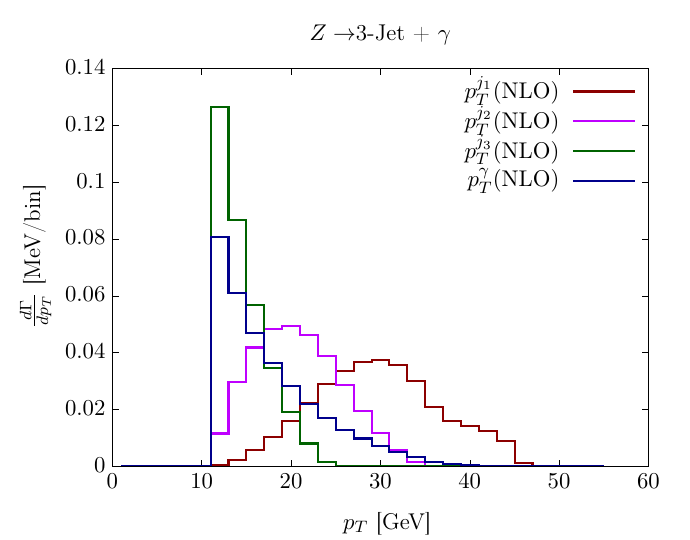}\\
              \hspace{0.8cm} (c)
			\caption{Distributions of the width with respect to $p_T$.}
			\label{fig:qqp_pT}
		\end{figure}

For the process $ Z \to q {\bar q} \gamma$, the same observables' distributions are plotted as that for the process $ Z \to q {\bar q}$. At the NLO level, we could have 1, 2, or 3-jet events, along with a photon. Again, the width for the 1-jet case is about an order of magnitude smaller than that for a 2-jet case. For this process, apart from jets, we have plotted observables for a photon and a photon-jet system. In Fig.~\ref{fig:qqp_pT}, we have plotted the $p_T$ distributions for each type of event. Here, for the 1-jet events, the $p_T$ of the jet is harder due to the presence of a photon. The NLO values for the width are smaller due to overall negative QCD corrections and the distribution of events over multiple jets. For 2 and 3-jet events, as the photon is emitted from the (anti-)quark, it is softer, and the distribution peaks near the $p_T$ cut value. As the jets are ordered according to their $p_T$ value, the hardest jet $p_T$ distribution peaked towards $M_Z/2$ value. Other jets are less hard.

\vspace{-0.3cm}

\vspace{-0.9cm}
\begin{figure}[H]
			\centering
			\includegraphics[width=0.36\linewidth]{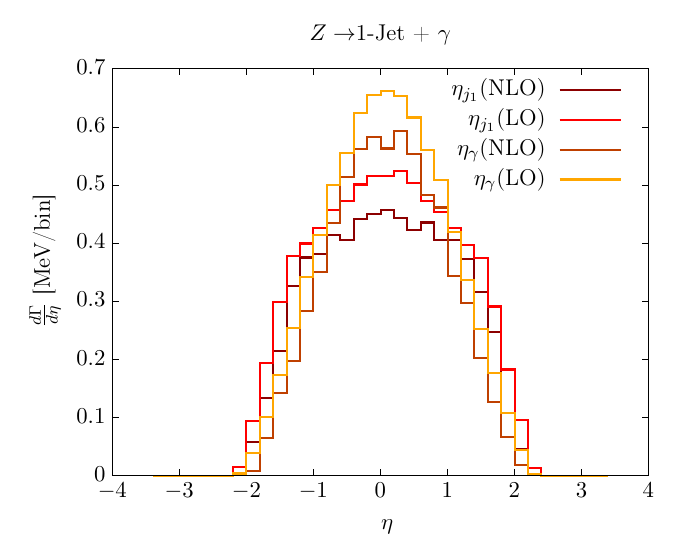}
            \includegraphics[width=0.36\linewidth]{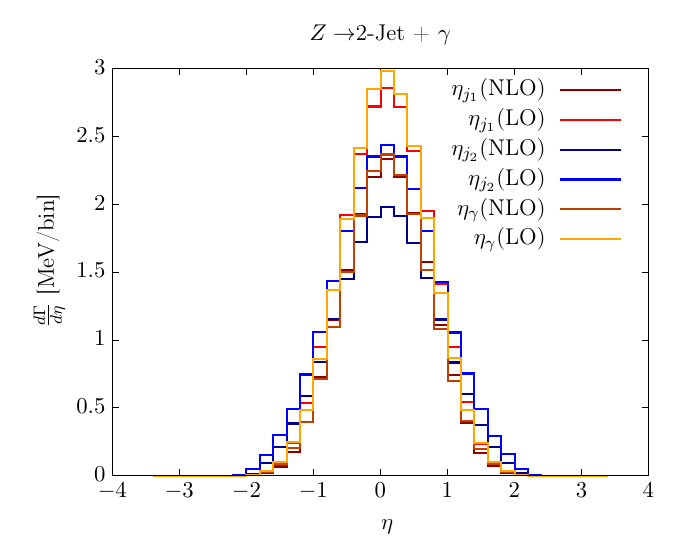}\\
			\hspace{1.0cm}(a)\hspace{5.4cm}(b)\\ 
            \includegraphics[width=0.36\linewidth]{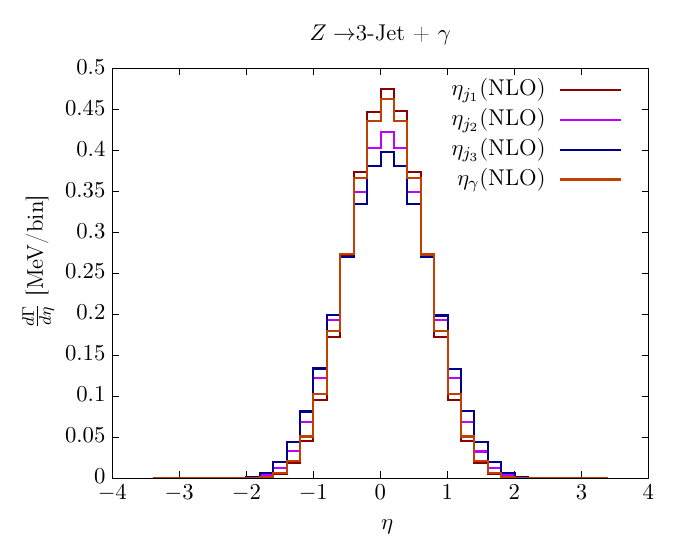}\\
            \hspace{0.8cm}(c)
	
			\caption{Distributions of the width with respect to $\eta$.}
			\label{fig:qqp_Eta}
		\end{figure}

\begin{figure}[H]
			\centering
			\includegraphics[width=0.40\linewidth]{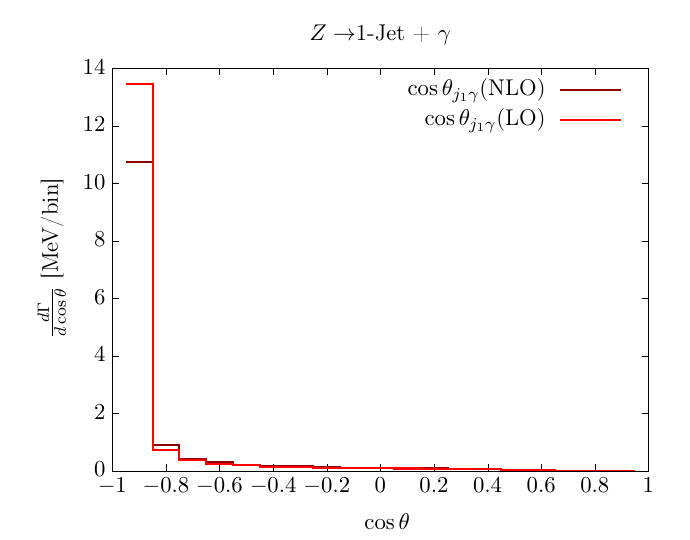}
            \includegraphics[width=0.40\linewidth]{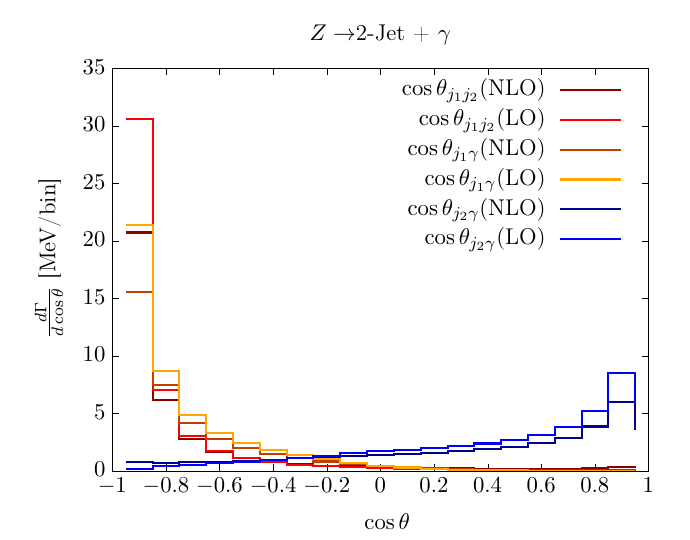}\\
			\hspace{1.0cm}(a)\hspace{6.0cm}(b)\\ 
            \includegraphics[width=0.40\linewidth]{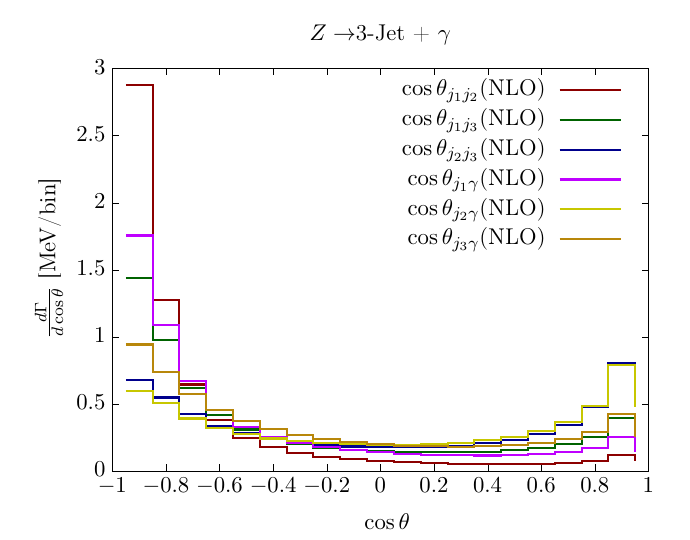}\\
            \hspace{0.9cm}(c)
	
			\caption{Distributions of the width with respect to  $\cos\theta$.}
			\label{fig:qqp_csT}
		\end{figure}

\begin{figure}[H]
			\centering
			\includegraphics[width=0.38\linewidth]{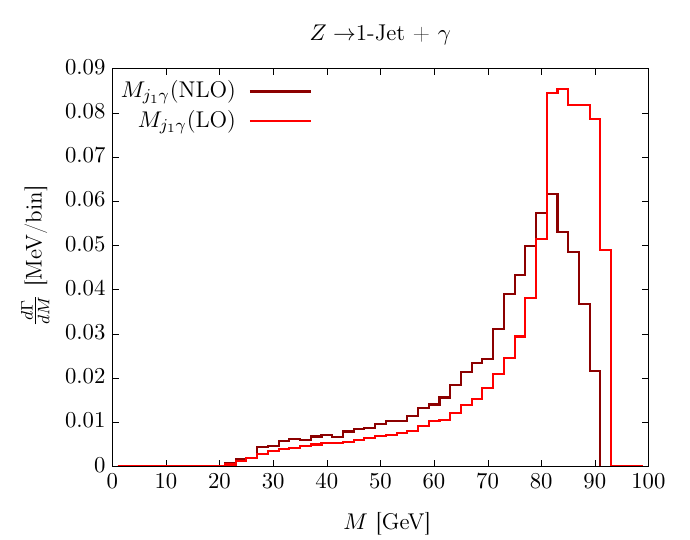}
            \includegraphics[width=0.38\linewidth]{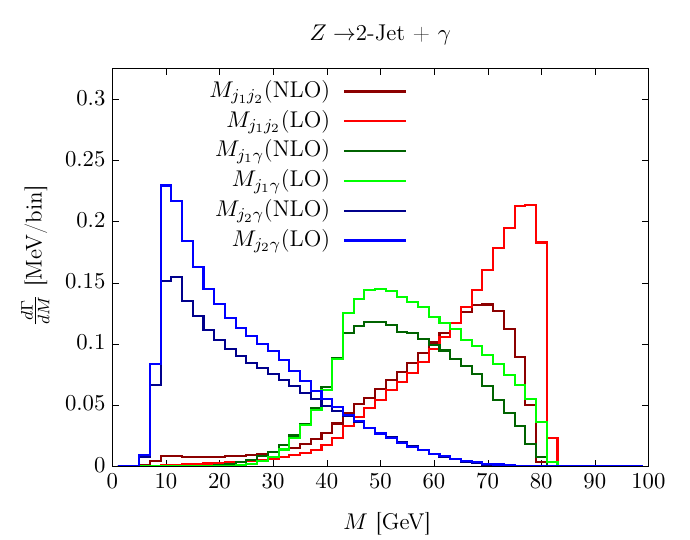}\\
			\hspace{1.0cm}(a)\hspace{5.7cm}(b)\\ 
	        \includegraphics[width=0.38\linewidth]{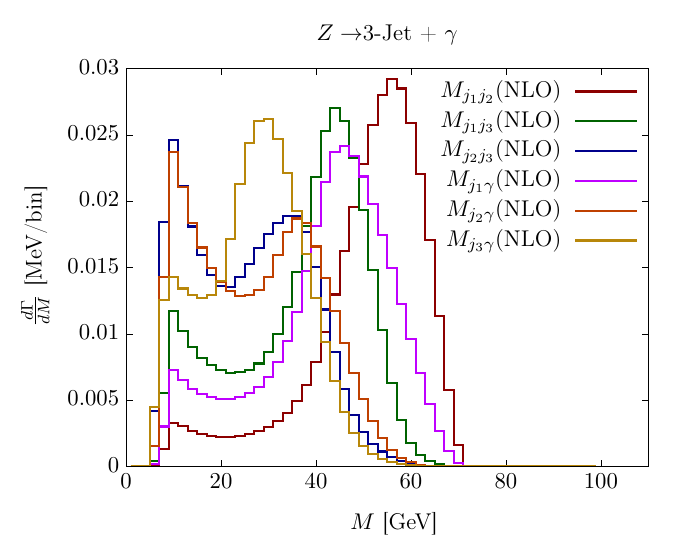}
            \includegraphics[width=0.38\linewidth]{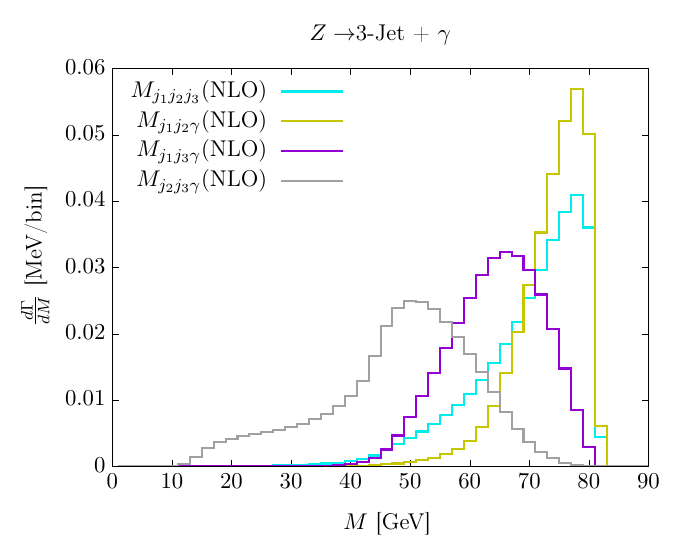}\\
            \hspace{1.0cm}(c)\hspace{5.7cm}(d)
			\caption{Distributions of the width with respect to to mass of a two-jet system.}
			\label{fig:qqp_mass}
		\end{figure}

In Fig.~\ref{fig:qqp_Eta}, we have plotted the pseudo-rapidity distributions.
For all jet scenarios, the distributions are peaked towards central values. The width of the distributions is becoming narrower as we go to a larger number of jet scenarios. In Fig.~\ref{fig:qqp_csT}, we have plotted the cosine of angles between jets and a jet and a photon.
For the 1-jet case, the jet and the photon are largely back-to-back. For the 2-jet case, the jets are largely back-to-back. The photon is closer to the jet it is emitted from. It makes a larger angle with the other jet. In Fig.~\ref{fig:qqp_mass}, the invariant masses of the jets and jet-photon systems are plotted. For the 1-jet case, the mass of the jet-photon system peaked towards the mass of the $Z$ boson. For the 2-jet case, the mass of the two-jet system, and that of the hardest-jet-photon system are distributed towards the mass of the $Z$ boson, while the mass of the photon and softer jets are peaked towards the lower value. For the 3-jet case, as expected, the mass of three particles peaked towards the mass of $Z$ boson, whereas for the mass of two particles, it is according to the jet $p_T$ distributions.

\begin{figure}[H]
			\centering
			\includegraphics[width=0.38\linewidth]{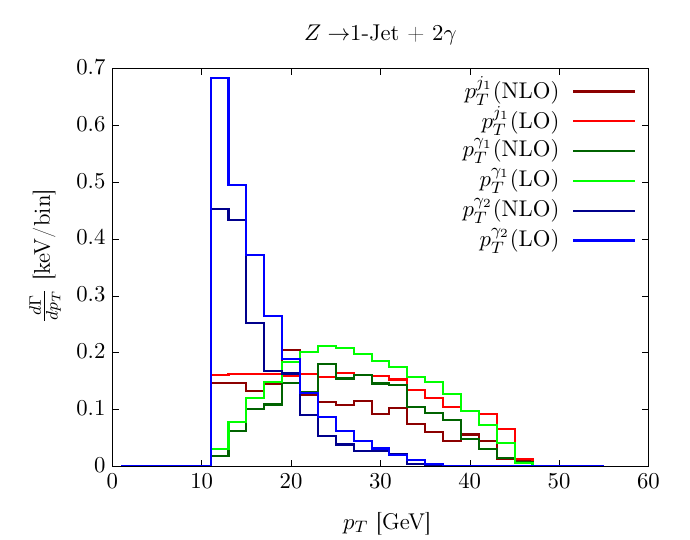}
            \includegraphics[width=0.38\linewidth]{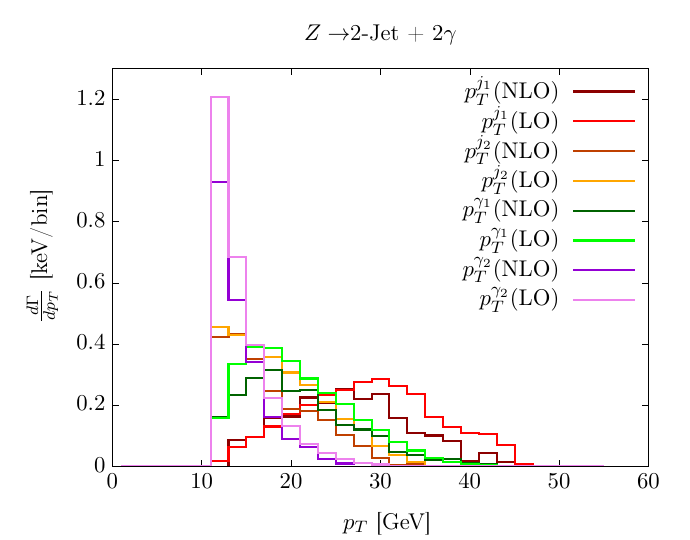}\\
            
			\hspace{1.0cm}(a)\hspace{5.7cm}(b)\\ 
	        \includegraphics[width=0.38\linewidth]{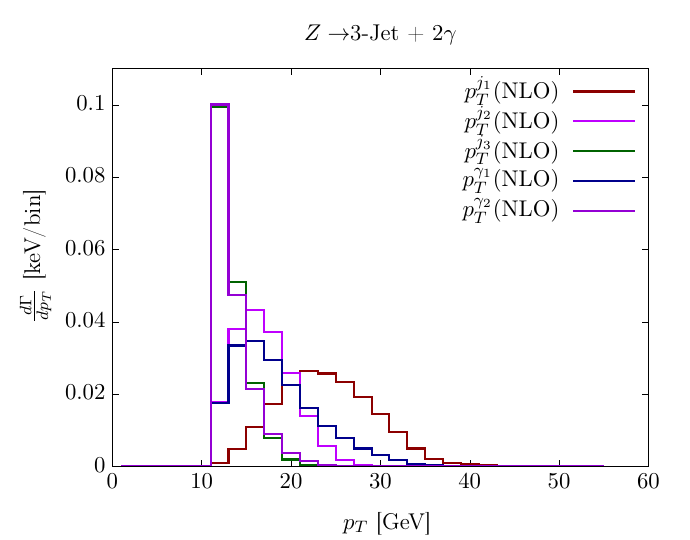}\\
            \hspace{0.9cm}(c)
			\caption{Distributions of the width with respect to  $p_T$.}
			\label{fig:qqpp_pT}
		\end{figure}

For the process $Z \to q {\bar q} \gamma \gamma$, we have plotted the distributions of similar observables as that for the process $Z \to q {\bar q} \gamma$. At the NLO level, we could have 1, 2, or 3-jet events, along with two photons. Again the width for the 1-jet case is about an order of magnitude smaller than that for a 2-jet case. Also, NLO distributions have smaller width values due to overall negative corrections and distribution of events over multiple jet scenarios. 
For this process, apart from jets, we have plotted observables for photons and photon-jet systems. Again the jets and photons are ordered according to their $p_T$, but separately. In Fig.~\ref{fig:qqpp_pT}, we have plotted the $p_T$ distributions of the $p_T$ ordered jets and photons. For the 1-jet events, as expected, for both the LO and the NLO levels, the $p_T$ distributions of the jet and one of the photons is hard. Another photon has a softer $p_T$ distribution. For the 2-jet and 3-jet cases, the situation is similar.

\vspace{-0.4cm}

\section{Conclusions}
\label{sec:conclusion}
We have computed the one-loop QCD corrections to the widths and decay distributions for the three processes (a) $Z\rightarrow q \bar{q}$, (b) $Z\rightarrow q \bar{q} \gamma$, and (c) $Z\rightarrow q \bar{q} \gamma \gamma$. The results for the process $Z\rightarrow q \bar{q}$ are in agreement with existing literature. There is an increase of $3.77\%$ in the partial width by one-loop QCD corrections to this process. At the NLO QCD level, there is a significant decrease in the widths of the processes $Z\rightarrow q \bar{q} \gamma$, and  $Z\rightarrow q \bar{q} \gamma \gamma$,  about $6.03\%$ and $12.39\%$ respectively. These processes will give us events with 1, 2, and 3-jet with one/two photons. The NLO corrections decrease the partial widths for the decay modes of 1, 2-jet with one/two photons. The NLO corrected partial widths for decay mode 1 and 2-jet with one photon are 1.38~MeV and 3.74~MeV, with the corresponding NLO corrections being $-13.75\%$ and $-20.26\%$, respectively. The corresponding values for the 1-jet and 2-jet cases with two photons are 3.26~keV and 4.39~keV, respectively, with NLO corrections of $-31.80\%$ and $-21.47\%$. Both of these decay channels, along with the NLO corrections, could be observed in future runs of the LHC or at future $e^{+}e^{-}$ colliders.

\bibliographystyle{JHEPCust.bst}
\bibliography{Z_Decay_QCD}
\vfill
\end{document}